\documentclass[a4paper,fleqn,usenatbib]{mnras}

\usepackage{newtxtext,newtxmath}
\usepackage[T1]{fontenc}
\usepackage{ae,aecompl}

\usepackage{graphicx}	% Including figure files
\usepackage{amsmath}	% Advanced maths commands
\usepackage{amssymb}	% Extra maths symbols

\usepackage{pdflscape}  % Rotate tables
\newcommand\kms{$\mbox{km s}^{-1}$}
\newcommand{\Teff}{\ensuremath{T_{\mathrm{eff}}}}%              % T_eff
\newcommand{\logg}{\ensuremath{\log g}}%                        % log g
\newcommand\pp{$\phantom{-}$}
\newcommand\z{$\phantom{0}$}
\title[Red Supergiants in NGC\,55]{Physical properties of the first spectroscopically confirmed red supergiant stars in the Sculptor Group galaxy NGC\,55}

\author[L.~R.~Patrick]{
L.~R.~Patrick,$^{1, 2, 3}$\thanks{E-mail: lpatrick@iac.es}
C.~J.~Evans,$^{3, 4}$
B.~Davies,$^{5}$
R-P.~Kudritzki,$^{6, 7}$
A.~M.~N. Ferguson,$^{3}$
\newauthor M.~Bergemann,$^{8}$
G.~Pietrzy{\'n}ski,$^{9, 10}$
and O. Turner$^{3}$
\\
$^{1}$Instituto de Astrof\'isica de Canarias, E-38205 La Laguna, Tenerife, Spain\\
$^{2}$Universidad de La Laguna, Dpto. Astrof\'isica, E-38206 La Laguna, Tenerife, Spain\\
$^{3}$Institute for Astronomy, University of Edinburgh, Royal Observatory Edinburgh, Blackford Hill, Edinburgh EH9 3HJ, UK\\
$^{4}$UK Astronomy Technology Centre, Royal Observatory Edinburgh, Blackford Hill, Edinburgh EH9 3HJ, UK\\
$^{5}$Astrophysics Research Institute, Liverpool John Moores University, Liverpool Science Park ic2, 146 Brownlow Hill, Liverpool L3 5RF, UK\\
$^{6}$Institute for Astronomy, University of Hawaii, 2680 Woodlawn Drive, Honolulu, HI, 96822, USA\\
$^{7}$University Observatory Munich, Scheinerstr. 1, D-81679, Munich, Germany\\
$^{8}$Max-Planck Institute for Astronomy, D-69117, Heidelberg, Germany\\
$^{9}$Universidad de Concepci{\'o}n, Departamento de Astronom{\'i}a, Casilla 160-C, Concepci{\'o}n, Chile\\
$^{10}$Nicolaus Copernicus Astronomical Centre, Polish Academy of Sciences, Bartycka 18, PL-00-716 Warszawa, Poland\\
}

\date{Accepted XXX. Received YYY; in original form ZZZ}

\pubyear{2016}

\begin{document}
\label{firstpage}
\pagerange{\pageref{firstpage}--\pageref{lastpage}}
\maketitle

\begin{abstract}
We present $K$-band Multi-Object Spectrograph (KMOS) observations of 18 Red
Supergiant (RSG) stars in the Sculptor Group galaxy NGC\,55.
Radial velocities are calculated and are shown to be in good
agreement with previous estimates,
confirming the supergiant nature of the targets and providing the first spectroscopically confirmed RSGs in NGC\,55.
Stellar parameters are estimated for 14 targets using the $J$-band analysis technique,
making use of state-of-the-art stellar model atmospheres.
The metallicities estimated confirm the low-metallicity nature of NGC\,55, in good agreement with previous studies.
This study provides an independent estimate of the metallicity gradient of NGC\,55, in excellent agreement with recent results published using hot massive stars.
In addition, we calculate luminosities of our targets and compare their distribution of effective temperatures and luminosities to other RSGs, in different environments, estimated using the same technique.

\end{abstract}


%%%%%%%%%%%%%%%%%%%%%%%%%%%%%%%%%%%%%%%%%%%%%%%%%%

%%%%%%%%%%%%%%%%% BODY OF PAPER %%%%%%%%%%%%%%%%%%

\section{Introduction} % (fold)
\label{sec:ngc55intro}

The Sculptor ``Group'' is a filament-shaped group of galaxies extending over
$\sim$5\,Mpc along the line of sight~\citep{1998AJ....116.2873J,2003A&A...404...93K},
containing three subgroups~\citep{2003A&A...404...93K} centred around NGC\,253
(d~$\sim$~4\,Mpc), NGC\,7793 (d~$\sim$~4\,Mpc) and NGC\,300 (d~$\sim$~2\,Mpc).
NGC\,55 is a member of the nearest subgroup that also contains NGC\,300 and a number of dwarf galaxies.
As one can resolve individual bright stars in these galaxies, the Sculptor Group provides us with a fantastic laboratory to investigate stellar and galactic evolution beyond the Local Group.

\begin{figure*}
 \centering
 \includegraphics[width=0.8\textwidth]{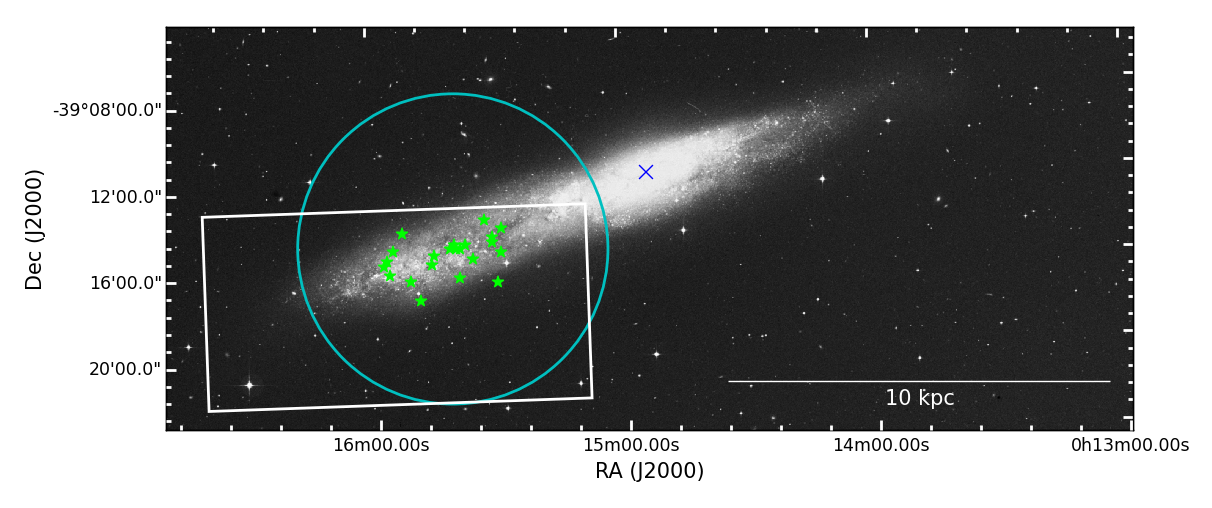}
 \caption{Digital sky survey image of NGC\,55 with the KMOS targets overlaid in green and the relevant footprint from the Araucaria Project
          \protect\citep{2006AJ....132.2556P} shown as a white rectangle.
          The 7$'$ diameter field of view of KMOS is indicated by the blue circle.
          The centre of NGC\,55, as defined by the Two Micron All Sky Survey
          \protect\citep[2MASS;][]{2006AJ....131.1163S},
          is shown as a blue cross.
         }
 \label{fig:ngc55-foot}
\end{figure*}

Table~\ref{tb:55prop} summarises the observational properties of NGC\,55,
which has an asymmetric and complicated morphology,
due to its high inclination
\cite[up to 80\degr;][]{1986A&A...166...97H,2013MNRAS.434.3511W}.
\cite{1961ApJ...133..405D} classified it as an LMC-like barred spiral galaxy (SB(s)m),
where the bar is seen along the line of sight,
prompting various claims that it is an edge-on analogue of the LMC.
Figure~\ref{fig:ngc55-foot} shows NGC\,55 and its complicated morphology,
where one can see the edge-on disk (along the major axis of the galaxy) and the brighter central part of the galaxy,
which represents the head of the bar.
In addition to NGC\,55 being orientated nearly edge on,
there are many star-formation features extending from the disk-bar system such as giant
\ion{H}{ii} regions and supergiant filaments and shells,
which are thought to allow ionising radiation to be transported to the halo
\citep{1996AJ....112.2567F}.

\begin{table}
\caption[NGC\,55 Basic Properties]{Basic properties of NGC\,55.\label{tb:55prop}}
% \scriptsize
\begin{center}
\begin{tabular}{lll}
  \hline
  \hline
Property & Value & Reference \\
  \hline
Classification & SB(s)m & {\cite{1991rc3..book.....D}}\\
$\alpha$ (J2000) & 00$^{\rm h}$ 14$^{\rm m}$ 53.602$^{\rm s}$ & {\protect\cite{2006AJ....131.1163S}}\\
$\delta$ (J2000) & $-$39 11 47.86 & {\protect\cite{2006AJ....131.1163S}}\\
Distance & 2.34\,$\pm$\,0.11\,Mpc & {\protect\cite{2016ApJ...829...70K}} \\
Inclination & 78\,$\pm$\,4\,\degr & {\protect\cite{1991AJ....101..447P},}\\
                                & & {\protect\cite{2013MNRAS.434.3511W}} \\
Radial velocity & 131\,$\pm$\,2\,\kms & {\protect\cite{2013MNRAS.434.3511W}}\\
\hline
\end{tabular}
\end{center}
\end{table}

The ionised gas morphology of NGC\,55, as well as its known population of massive hot stars
\citep{2008A&A...485...41C,2012A&A...542A..79C}, points to a recent history of intense star formation.
The infrared luminosity of NGC\,55 implies a star-formation rate of 0.22\,M$_{\odot}$yr$^{-1}$
\citep{2004ApJS..154..248E} and is dominated by a bright central region and disk emission,
which supports the hypothesis of recent star formation.
Here we present the first spectroscopic study of red supergiant stars (RSGs) in NGC\,55, with the objective of providing an independent check on the metallicity estimates from blue supergiants (BSGs) and \ion{H}{ii} regions.

From galaxy mass-metallicity relations, the metal content of NGC\,55 was expected to be LMC-like,
which is supported by
\cite{2012A&A...542A..79C}, who measured metallicities of 12 BSGs using optical spectroscopy and found a mean metallicity [Z]~=~$\log$(Z/Z$_{\odot}$)~=~$-$0.40\,$\pm$\,0.13\,dex.
In addition, the literature contains several metallicity measurements from \ion{H}{ii} regions across the disk of NGC\,55, finding a large spread of metallicities
\citep[$-$0.6 to $-$0.2 -- assuming a solar-like $\alpha$/Fe;][]{1983MNRAS.204..743W,1986A&A...154..352S,1994ApJ...420...87Z,2003A&A...412...69T,2012A&A...542A..79C}

In addition, the BSG work of
\cite{2012A&A...542A..79C} has recently been taken further by
\cite{2016ApJ...829...70K}, who analysed $\sim$60 BSGs across a large spatial extent of NGC\,55.
These authors found a central metallicity of
[Z]~=~$-0.37\,\pm\,0.03$\,dex, in good agreement with~\cite{2012A&A...542A..79C}.
Interestingly, these authors also found evidence for a metallicity gradient of $-0.22\,\pm\,0.06$\,dex/R$_{25}$, the first such evidence in this galaxy.
On the other hand,~\cite{2017MNRAS.464..739M} estimate abundances from 25 \ion{H}{ii} regions within NGC\,55 finding no evidence for an abundance gradient.
Therefore, it is important that we obtain independent estimates of metallicities in NGC\,55 to provide insight into this potential discrepancy.

Even though the hot massive-star population of NGC\,55 has been explored,
there are currently no confirmed RSGs in NGC\,55, although~\cite{2005ApJ...622..279D} noted that the near-IR colour--magnitude diagram (CMD) of fields within the disk reveal RSG candidates.
The current study represents the first quantitative analysis of RSGs in NGC\,55 and it will provide an independent estimate of the metallicity of the young stellar population within this galaxy.

In recent years, moderate resolution spectroscopy has been shown to be a
useful tool to determine the physical parameters of RSGs
\citep{2010MNRAS.407.1203D,2015ApJ...806...21D}.
In this context, the $K$-band Multi-Object Spectrograph
\cite[KMOS;][]{2013Msngr.151...21S} on the Very Large Telescope (VLT),
has been used to explore the chemical evolution of external galaxies using RSGs as cosmic abundances probes.
This analysis makes use of state-of-the-art stellar model atmospheres including
departures from the assumption of local thermodynamic equilibrium (LTE)
for the strongest absorption lines in a spectral window in the $J$-band
\citep{2012ApJ...751..156B,2013ApJ...764..115B,2015ApJ...804..113B}.
The first application of this technique with KMOS was by
\cite{2015ApJ...803...14P} who investigated RSGs in NGC\,6822 (d$\sim$0.5\,Mpc).
Building on this initial study we have explored the RSG population in several external galaxies within and beyond the Local Group of galaxies.

% Near-IR spectroscopy in the $J$-band (at $R$\,$\sim$\,3000) has been developed over recent years as a method to determine the physical parameters of RSGs
% \citep{2010MNRAS.407.1203D,2015ApJ...806...21D}.
% When combined with the $K$-band Multi-Object Spectrograph~\cite[KMOS;][]{2013Msngr.151...21S} on the Very Large Telescope (VLT), quantitative spectroscopy of samples of extragalactic RSGs is now feasible compared to past Galactic studies using high-resolution spectroscopy
% ~\citep[e.g.][]{Cunha07,Davies09a,Davies09b}.
% The first application of this technique with KMOS was by~\cite{2015ApJ...803...14P} who investigated RSGs
% in NGC\,6822 (d$\sim$0.5\,Mpc).

In addition to applying this technique to individual RSGs, we can estimate metallicities of more distant star clusters, in which the integrated near-IR flux is dominated by the contribution from RSGs~\citep{2013MNRAS.430L..35G}.
\cite{2015ApJ...812..160L} applied this technique to three super star clusters in NGC\,4038, one of the Antennae galaxies.
Further ahead, with a near-IR multi-object spectrograph on a 30-m class telescope,~\cite{2011A&A...527A..50E} demonstrated that individual metallicities of RSGs can be estimated at distances of tens of Mpc and, using the integrated-light analysis, at distances of hundreds of Mpc with clusters.

In this paper we present the analysis of KMOS observations of RSGs in NGC\,55.
Sections~\ref{sec:ngc55obs} and~\ref{sec:ngc55:obs_data} detail the target selection, observations and data reduction.
In Section~\ref{sec:ngc55results} the main results of the paper are presented and discussed,
which include radial-velocity estimates
-- confirming their membership of NGC\,55, and by inference nature as supergiants --
and the estimates of the physical parameters of our targets.
The conclusions of the study are summarised in Section~\ref{sec:ngc55conc}.

% section introduction (end)

\section{Target Selection} % (fold)
\label{sec:ngc55obs}

Targets were selected for spectroscopy based on their optical photometry from the Araucaria Project
\citep{2006AJ....132.2556P}, where the footprint of the relevant CCD is
highlighted with a large white rectangle in Figure~\ref{fig:ngc55-foot}.
This is ground-based data taken with the 1.3\,m Warsaw Telescope where the
pixel scale was 0\farcs25 and the typical seeing conditions were 1\farcs0.

\begin{figure}
  \centering
  \includegraphics[width=\columnwidth]{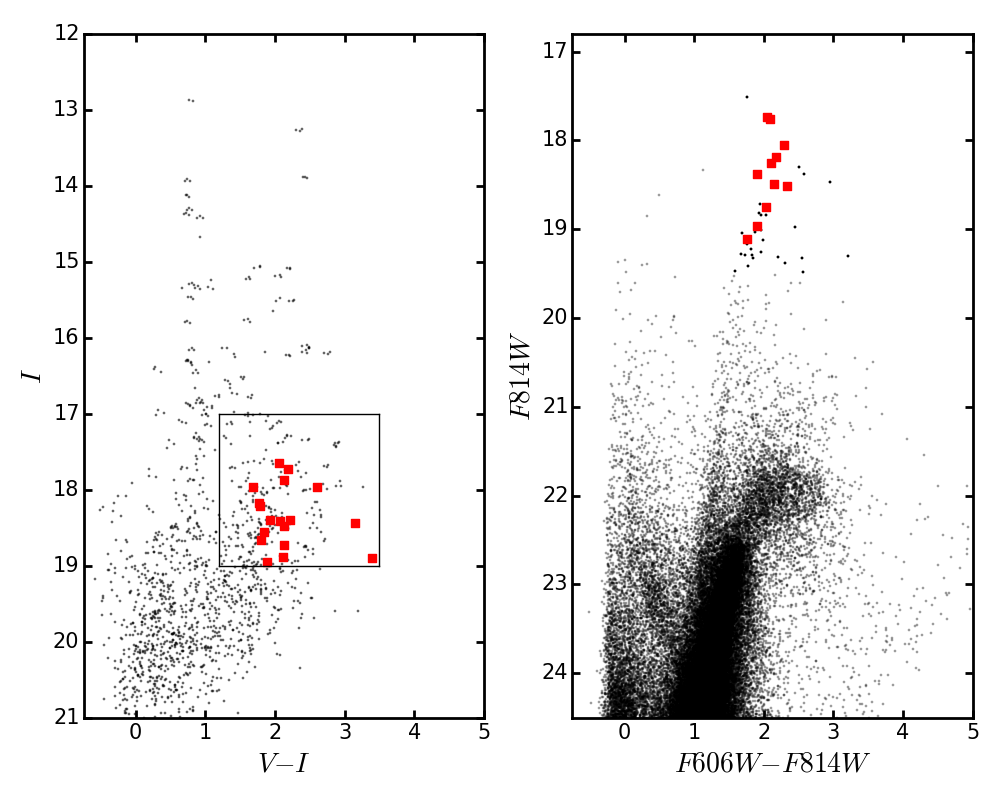}
 \caption[NGC\,55 ground- and space-based colour--magnitude diagrams]{
  Colour-magnitude diagrams used for target selection. Left-hand panel:
  ground-based photometry from~\citet{2006AJ....132.2556P} for the area highlighted in Figure~\ref{fig:ngc55-foot}.
  The black box was used to select candidate RSGs (17\,$<$\,$I$\,<19, 1.2\,$<$\,$V-I$\,$<$\,3.5).
  Right-hand panel: overlapping $HST$ photometry of part of the KMOS field from~\cite{2009ApJS..183...67D}.
  Observed targets are highlighted with red squares.
          }
 \label{fig:VI}
\end{figure}

The optical CMD used to select targets is displayed in the left-hand panel of
Figure~\ref{fig:VI}, where the RSG candidates are within the black box and the
observed targets are highlighted in red.

Candidate RSGs are defined as having $17 < I < 19$ and $1.2 < V-I < 3.5$.
% The magnitude limit here is a result of the expected magnitudes of RSGs in NGC\,55.
The magnitude limits were chosen to give S/N$\sim$100 for the requested exposure times of the observations~\citep[cf.][which used similar selection criteria for NGC\,300]{2015ApJ...805..182G},
however the requested programme was not fully realised and hence the final S/N of the spectra is somewhat lower.
The $V-I$ colour limit is chosen to avoid potential Galactic contamination at $V-I\sim$ 0.95.
All stars that meet these criteria are defined as candidate RSGs and the final target selection was performed by analysing the spatial location of the candidates and selecting those which traced the disk population of NGC\,55 (see Figure~\ref{fig:ngc55-foot}).

In addition to the ground-based photometry available, the ACS Nearby Galaxy Survey Treasury
\citep[ANGST;][]{2009ApJS..183...67D} project has publicly-available photometry for several fields within the disk of NGC\,55,
although we preferred use of the ground-based data for its more comprehensive coverage in the south-east part of the NGC\,55 disk. A $F814W$, $F606W-F814W$ CMD is shown in the right-hand panel of
Figure~\ref{fig:VI} for the ACS ANGST field which contains 11 of our KMOS targets, highlighted in red.
Given the smaller spatial coverage of the $HST$ ACS field, the expected number of Galactic contaminants is corresponding smaller.
As a result of their cool temperatures and extreme luminosities, RSGs should exist in a ``plume'' at the tip of a structure of cool stars in the $F606W-F814W$, $F814W$ CMD.
This compares well to the location (in colour--magnitude parameter space) of confirmed RSGs in NGC\,300~\citep{2015ApJ...805..182G}.

% subsection target_selection (end)
\section{Observations and Data Reduction} % (fold)
\label{sec:ngc55:obs_data}

The data were obtained as part of the KMOS Guaranteed Time Observations (GTO, 092.B-0088) to investigate stellar metallicities in NGC\,300 and NGC\,55.
We observed one KMOS configuration in NGC\,55, located in the south-eastern part of the main disk (as shown in Figure~\ref{fig:ngc55-foot}).
Initial observations were obtained in 2013 October, with additional data taken in 2014 September as part of back-up observations during a different part of the GTO programme.
Unfortunately the conditions in the 2014 run did not lead to improvements on the 2013 data, and these observations are not considered further here.

The observations consisted of a total of 20\,$\times$\,600\,s integrations, where object exposures (O) and sky-offset exposures (S) interleaved in a O, S, O pattern;
taken over the nights of 2013 October 14 and 15,
in seeing conditions of 0\farcs8 to 1\farcs2.
In addition, on each night a standard set of KMOS calibration files was obtained as well as observations of a standard star for telluric correction.
HIP\,3820
\citep[B8\,V;][]{1978mcts.book.....H} was observed using the 24-arm telluric template (KMOS\_spec\_acq\_stdstarscipatt).
However, on 2013 October 14, this OB was interrupted and could not be repeated owing to operational issues, hence several of the IFUs were not observed with the 24-arm template.

% In September 2014 only the 3-arm telluric template (KMOS\_spec\_cal\_stdstar) was observed as opposed to the full 24-arm template. However, on both of these nights both HIP\,3820 and HIP\,18926~\citep[B3\,V;][]{1988mcts.book.....H} were observed as standard stars.

% The quality of the observations taken on each night varies significantly.
% The first set of observations (October 2013) were taken in excellent conditions where the seeing was stable and with good transparency.
% As one would expect with back-up observations, the conditions were not so idyllic,
% with large variations in seeing and with patchy, sometimes significant, cloud coverage.

Differences in the quality of the data and in the actual execution of the observations must all be taken into account when the data is reduced.
% In addition to the varying observing conditions of the data, there are also changes in the spectral resolution of the instrument due to observations at different rotator angles on the Nasmyth axis.
The delivered spectral resolutions of our data (from analysis of the relevant arc calibration frames) are summarised in Table
\ref{tb:55res}, where changes in the spectral resolution are principally the result of differences in rotator angles of the Nasmyth axis.
These were taken into account when combining exposures from different nights by degrading the spectra to the lowest effective resolving power
(using a simple Gaussian filter).
For example, spectra for the target in IFU~1 were degraded to a resolving power of 3300 before combining into a master spectrum (the resolution of the Ar\,$\lambda$1.21430\,$\mu$m line is chosen as this more accurately represents the spectral resolution where the diagnostic lines are present).

\begin{table*}
\caption[Measured velocity resolution for each night]
{Measured velocity resolution and resolving power across each detector.\label{tb:55res}}
% \scriptsize
\begin{center}
\begin{tabular}{ccrcccc}
\hline
\hline
Date & Spectrograph & IFUs & \multicolumn{2}{c}{Ne\,$\lambda$1.17700\,$\mu$m}
            & \multicolumn{2}{c}{Ar\,$\lambda$1.21430\,$\mu$m} \\
& & & FWHM (\kms) & $R$ & FWHM (\kms) & $R$ \\
  \hline
  \\
           & 1 & 1-8 &   95.48\,$\pm$\,2.42 & 3140\,$\pm$\,80 &
                         90.71\,$\pm$\,2.09 & 3305\,$\pm$\,76 \\
14-10-2013 & 2 & 9-16 &  88.67\,$\pm$\,1.67 & 3381\,$\pm$\,64 &
                         86.35\,$\pm$\,1.84 & 3472\,$\pm$\,74 \\
           & 3 & 17-24 & 82.89\,$\pm$\,1.81 & 3617\,$\pm$\,79 &
                         80.56\,$\pm$\,2.11 & 3721\,$\pm$\,97 \\
                         \\
  \hline
  \\
           & 1 & 1-8 &   95.48\,$\pm$\,2.46 & 3140\,$\pm$\,81 &
                         90.78\,$\pm$\,2.12 & 3302\,$\pm$\,77 \\
15-10-2013 & 2 & 9-16 &  88.91\,$\pm$\,1.66 & 3371\,$\pm$\,63 &
                         86.30\,$\pm$\,1.85 & 3473\,$\pm$\,74 \\
           & 3 & 17-24 & 82.96\,$\pm$\,2.14 & 3612\,$\pm$\,76 &
                         80.77\,$\pm$\,2.14 & 3712\,$\pm$\,98 \\
                         \\
% \hline
% \\
%            & 1 & 1-8 &   84.18\,$\pm$\,1.93 & 3561\,$\pm$\,82 &
%                          81.76\,$\pm$\,2.15 & 3667\,$\pm$\,96 \\
% 14-09-2014 & 2 & 9-16 &  87.00\,$\pm$\,1.69 & 3446\,$\pm$\,67 &
%                          84.67\,$\pm$\,1.93 & 3541\,$\pm$\,81 \\
%            & 3 & 17-24 & 97.14\,$\pm$\,1.88 & 3086\,$\pm$\,60 &
%                          94.85\,$\pm$\,2.01 & 3161\,$\pm$\,67 \\
%                          \\

% \hline
% \\
%            & 1 & 1-8 &   82.55\,$\pm$\,1.96 & 3632\,$\pm$\,86 &
%                          80.41\,$\pm$\,2.30 & 3728\,$\pm$\,106\\
% 15-09-2014 & 2 & 9-16 &  88.08\,$\pm$\,1.78 & 3404\,$\pm$\,69 &
%                          86.03\,$\pm$\,1.96 & 3485\,$\pm$\,80\z\\
%            & 3 & 17-24 & 98.04\,$\pm$\,1.91 & 3058\,$\pm$\,59 &
%                          96.74\,$\pm$\,2.05 & 3099\,$\pm$\,66\z\\
%                          \\
\hline
\end{tabular}
\end{center}
\end{table*}

The observations were reduced using the recipes provided by the Software Package for Astronomical Reduction with KMOS
\citep[SPARK;][]{2013A&A...558A..56D}.
The standard KMOS/esorex routines were used to calibrate and reconstruct the science and standard-star data cubes as outlined by
\cite{2013A&A...558A..56D}, including correction for the readout column bias as well as enhancing the bad-pixel mask following Turner et al. (in prep.).
Using the reconstructed data cubes the pipeline was used to extract science and sky spectra in a self-consistent way for all exposures.

Sky subtraction was performed using the ESO {\sc skycorr} package~\citep{2014A&A...567A..25N}.
{\sc skycorr} is an instrument independent tool that applies a scaling to a sky spectrum given a pair of target and sky spectra in order to more accurately match the sky lines in the target spectrum and hence, provide a more optimised sky subtraction.
This works by adapting the reference sky spectrum to correct for differences as a result of temporal and spatial airglow variability.
This software is specifically designed for observations at Cerro Paranal and has been shown to be an effective tool for various science goals~\citep[e.g.][]{2014A&A...567A..25N,2015ApJ...805..182G,2016MNRAS.455.2028F,2016MNRAS.457.1468L}.

Telluric correction is performed on each sky-subtracted spectrum (before combination) using the method described in full by~\cite{2015ApJ...803...14P}.
Briefly, additional corrections are made to the standard KMOS/esorex method of telluric correction by correcting for potential offsets between the wavelength solutions of the science and telluric spectra using a iterative cross-correlation approach.
In addition, a simple scaling is applied to the telluric spectrum in order to more accurately match the telluric absorption in the science spectrum.
Once these additional corrections have been implemented,
the science spectra are divided through by the telluric spectrum to remove the telluric contamination.

To combine the fully calibrated and corrected spectra from each night,
the spectra are degraded to the lowest resolving power measured for their respective IFUs.
Once each set of spectra have a constant resolution, any differences in the wavelength
solution are corrected using an iterative cross-correlation approach where each
spectrum is corrected to the wavelength solution of a single reference spectrum, for each target.
The choice of the reference spectrum for each target is important as, if a
poor-quality spectrum (with strong residuals from the sky or telluric correction)
were selected, there could be an alignment of the residuals leading to a spurious signal when combined.
To avoid this, the highest-quality spectrum for each target is selected as the reference spectrum.
In practise, this was typically the fourth exposure from the night of 2013 October 15.

Over an individual night there were generally not significant shifts, so this procedure mainly corrects for differences arising between each night.
Once this correction is implemented all spectra were combined using a simple median combine.
This simple method is preferred to something more sophisticated as there were significant sky- and telluric-correction residuals present in many of these spectra, the impact of which was minimised when taking the median of the data.

% section observations_and_data_reduction (end)
% section observations (end)

\section{Results and Discussion} % (fold)
\label{sec:ngc55results}

\subsection{Radial Velocities} % (fold)
\label{sub:rvs}

To measure radial velocities, the accuracy of the wavelength solution is vital.
For this reason -- for the radial-velocity analysis --
the data is reduced using a slightly different approach to the one outlined in Section
\ref{sec:ngc55:obs_data}.
Here, the KMOS/esorex pipeline is used to perform the sky subtraction and combine the frames.
After this step, the accuracy of the wavelength solution provided by the KMOS
pipeline for each target spectrum (before telluric correction)
is checked with a reference telluric spectrum using an iterative cross-correlation approach.

After correction for any shifts in the telluric lines, an initial (stellar) radial velocity was estimated via cross-correlation with a model stellar spectrum over the 1.16-1.21\,$\mu$m region.
Using this estimate as an initial guess, radial velocities are independently measured for several strong spectral features within this region.
The final radial velocity is an average of the measurements made using the individual spectral features, where the dispersion defines the uncertainty on the measurement.
This method has been shown to work well on stellar spectra from KMOS
\citep{2015ApJ...798...23L,2015ApJ...803...14P,2016MNRAS.458.3968P}.
This method is preferred to the result using the cross-correlation of the whole
1.16--1.21\,$\mu$m region as residual reduction features can act to perturb this estimate on large scales,
whereas this is more easily identified (by eye) using smaller,
well-defined spectral features.
Heliocentric corrections were calculated using ESO's Airmass calculator.
\footnote{https://www.eso.org/sci/observing/tools/calendar/airmass.html}

Estimated radial velocities for each night are listed in
Table~\ref{tb:n55obs-params} and an average is taken for each target.
Uncertainties quoted on the average are the standard deviation of the measurements on each night.
Three targets have been excluded from this radial velocity analysis analysis (and are not listed in Table~\ref{tb:n55obs-params}) based on their spectral
appearance and unreliable, inconsistent radial-velocity estimates
-- owing to reduction residuals.
% Therefore, henceforth, these targets are excluded from the sample.

\begin{figure}
 \centering
 \includegraphics[width=\columnwidth]{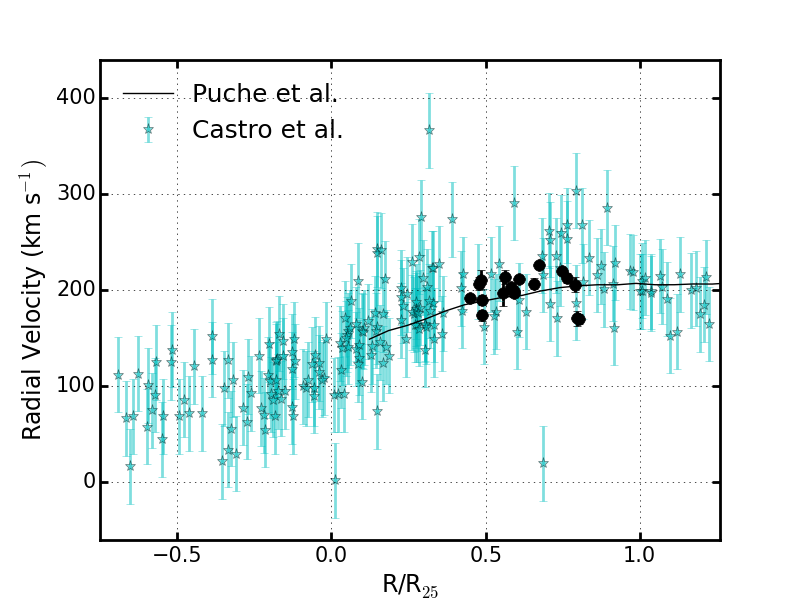}
 \caption[Radial velocities for KMOS targets shown against projected radius]{
 Radial velocities for the KMOS RSGs (black points) shown against projected radius from the centre of NGC\,55 as defined by the Two Micron All Sky Survey~\citep[2MASS;][]{2006AJ....131.1163S} scaled by R$_{25}$~=~16.2\,$\pm$\,0.4\,arcmin~\citep{1991rc3..book.....D}.
Blue stars show data for $\sim$200 BSGs in NGC\,55 from~\citet[][shown with 50\% transparency to highlight densely populated areas]{2008A&A...485...41C} alongside the rotation curve of NGC\,55~\citep[black solid line;][]{1991AJ....101..447P}.}
 \label{fig:RvsRV}
\end{figure}

Comparing the estimated velocities to previous measurements we find good agreement with velocities measured for $\sim$200 BSGs in NGC\,55 by
\cite{2008A&A...485...41C}, as well as with measurements of the velocity from the \ion{H}{i} gas~\citep{1991AJ....101..447P}.
The estimated radial velocities as a function of the de-projected galactocentric distance are shown in Figure~\ref{fig:RvsRV}, where previous measurements are also shown for comparison.
The radius at which the surface brightness first reaches 25\,mag/arcsec$^2$ in the $B$-band is shown for scale~\citep[R$_{25}$~=~16.2\,$\pm$\,0.4\,arcmin][]{1991rc3..book.....D}.
We find no evidence for a systematic offset between the measurements of~\cite{2008A&A...485...41C} and those measured in this study.
The de-projected radius is calculated by assuming the geometrical model defined by~\citet{1991AJ....101..447P} and updated by~\citet{2013MNRAS.434.3511W}, i.e. with an inclination angle $i$~=~78\,$\pm$\,4\,\degr and a position angle $\phi$~=~109\,\degr.

% subsection radial_velocities (end)
\subsection{Stellar Parameters} % (fold)
\label{sub:stellar_parameters}

Stellar parameters are estimated for each target using the $J$-band analysis technique
(given in Table~\ref{tb:ngc55fit-pars}).
This technique is described and tested in detail by
\cite{2010MNRAS.407.1203D,2015ApJ...806...21D} and applied to KMOS data by
\cite{2015ApJ...803...14P,2015ApJ...805..182G,2015ApJ...812..160L} and
\cite{2016MNRAS.458.3968P}.
This analysis technique uses a grid of synthetic spectra extracted from stellar model atmospheres to estimate overall metallicity ([Z]~=~$\log$(Z/Z$_{\odot}$)), effective temperature (\Teff), surface gravity (\logg) and microturbulent velocity ($\xi$).
This technique is tailored specifically to the resolution of KMOS spectra, and with a S/N~$\ge~100$ metallicities can be estimated to a precision of $\pm$\,0.10\,dex.
The range in the parameters used are applicable for RSGs and span
$-1.0~\le~$[Z]$~\le~1.0$,
$3400~\le~\Teff~\le~4400$\,K,
$-1.00~\le~\logg~\le~1.00$ and
$1.0~\le~\xi~\le~5.0$\,\kms.

\begin{figure*}
 %\vspace{302pt}
 \begin{center}
 \centering
\includegraphics[width=0.75\textwidth]{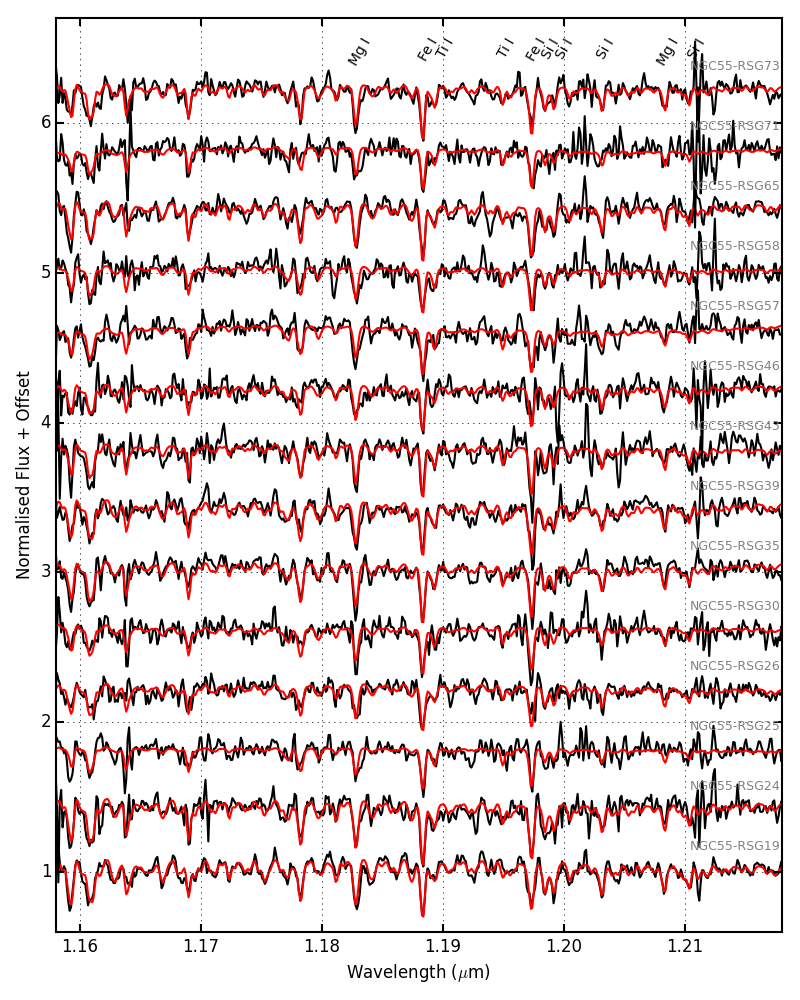}
\caption[Observed and best-fit model spectra of RSGs in NGC\,55]{Observed and best-fit model spectra of RSGs in NGC\,55 (black and red lines, respectively).
The lines used for the analysis, from left-to-right by species, are
Fe\,{\scriptsize I}$\,\lambda\lambda$1.188285,
1.197305;
Mg\,{\scriptsize I}$\,\lambda\lambda$1.182819,
1.208335;
Si\,{\scriptsize I}$\,\lambda\lambda$1.198419,
1.199157,
1.203151,
1.210353;
Ti\,{\scriptsize I}$\,\lambda\lambda$1.189289,
1.194954.\label{fig:ngc55spec}}
\end{center}
\end{figure*}

Observations are compared with synthetic RSG spectra,
extracted from {\sc marcs} stellar model atmospheres
\citep{2008A&A...486..951G}, with non-LTE
corrections computed for the strongest iron, titanium, silicon and magnesium absorption lines
\citep{2012ApJ...751..156B,2013ApJ...764..115B,2015ApJ...804..113B}.
The synthetic spectra are compared with observations using the reduced $\chi$-squared statistic for the 10 diagnostic absorption lines marked in Figure~\ref{fig:ngc55spec}.
Best-fit parameters are assessed using a sampling of the posterior probability density function using {\sc emcee}~\citep{2013PASP..125..306F}.

\begin{table*}
\begin{center}
\caption{Physical parameters determined for the KMOS targets in NGC\,55.\label{tb:ngc55fit-pars}}
% \scriptsize
\begin{tabular}{lr ccccc}
 \hline
 \hline
  Target  & IFU & S/N & $\xi$ (\kms) & [Z] & log\,$g$ & \Teff (K)\\
  \hline
NGC55-RSG19 & 6  & 35 & 3.9\,$\pm$\,0.7 & $-$0.23\,$\pm$\,0.22 & $-$0.8\,$^{+1.2}_{-0.3}$ & 4260\,$^{+80}_{-210}$\\
NGC55-RSG24 & 10 & 30 & 4.2\,$\pm$\,0.8 & $-$0.43\,$\pm$\,0.32 & $-$0.4\,$^{+1.0}_{-0.4}$ & 3920\,$^{+330}_{-370}$\\
NGC55-RSG25 & 8  & 30 & 3.3\,$\pm$\,1.1 & $-$0.65\,$\pm$\,0.32 & \pp0.3\,$^{+0.3}_{-1.0}$ & 3480\,$^{+570}_{-30}$\\
NGC55-RSG26 & 4  & 25 & 3.1\,$\pm$\,1.0 & $-$0.59\,$\pm$\,0.32 & $-$0.4\,$^{+0.7}_{-0.5}$ & 3890\,$^{+280}_{-320}$\\
NGC55-RSG30 & 3  & 25 & 3.6\,$\pm$\,1.0 & $-$0.49\,$\pm$\,0.32 & \pp0.4\,$^{+0.2}_{-1.0}$ & 3660\,$^{+360}_{-180}$\\
NGC55-RSG35 & 12 & 30 & 4.2\,$\pm$\,0.6 & $-$0.50\,$\pm$\,0.30 & $-$0.2\,$^{+0.5}_{-0.7}$ & 3810\,$^{+310}_{-250}$\\
NGC55-RSG39 & 14 & 30 & 3.1\,$\pm$\,0.9 & $-$0.27\,$\pm$\,0.26 & $-$0.8\,$^{+1.1}_{-0.1}$ & 3830\,$^{+220}_{-270}$\\
NGC55-RSG43 & 24 & 25 & 3.5\,$\pm$\,0.9 & $-$0.40\,$\pm$\,0.29 & \pp0.1\,$^{+0.3}_{-0.9}$ & 3680\,$^{+390}_{-160}$\\
NGC55-RSG46 & 22 & 30 & 2.7\,$\pm$\,0.9 & $-$0.50\,$\pm$\,0.32 & \pp0.4\,$^{+0.2}_{-1.0}$ & 3950\,$^{+300}_{-490}$\\
NGC55-RSG57 & 1  & 35 & 3.8\,$\pm$\,0.8 & $-$0.56\,$\pm$\,0.32 & \pp0.1\,$^{+0.3}_{-0.3}$ & 3540\,$^{+150}_{-90}$\\
NGC55-RSG58 & 15 & 30 & 3.2\,$\pm$\,1.0 & $-$0.46\,$\pm$\,0.31 & \pp0.1\,$^{+0.3}_{-0.9}$ & 3610\,$^{+320}_{-150}$\\
NGC55-RSG65 & 17 & 30 & 3.5\,$\pm$\,0.9 & $-$0.54\,$\pm$\,0.31 & $-$0.6\,$^{+0.8}_{-0.3}$ & 3910\,$^{+330}_{-360}$\\
NGC55-RSG71 & 20 & 25 & 2.6\,$\pm$\,1.1 & $-$0.71\,$\pm$\,0.28 & \pp0.4\,$^{+0.3}_{-1.0}$ & 3560\,$^{+230}_{-120}$\\
NGC55-RSG73 & 19 & 25 & 3.5\,$\pm$\,0.9 & $-$0.38\,$\pm$\,0.30 & \pp0.2\,$^{+0.4}_{-0.9}$ & 3920\,$^{+270}_{-340}$\\

  \hline
  \end{tabular}
  \end{center}
\end{table*}

Prior assumptions are implemented based on the luminosity of each target,
combined with an acceptable range in mass, based on their spectroscopic
confirmation as RSGs. Luminosities have been calculated with ground-based
$I$-band photometry using the bolometric correction of
\citet{2013ApJ...767....3D}, where extinction is accounted for assuming
$E(B-V)$~=~0.15 and a distance of 2.34\,Mpc~\citep{2016ApJ...829...70K}.
At this point, it is important to note that the bolometric correction of
\citet{2013ApJ...767....3D} is calibrated for a fixed temperature.
This may introduce a small uncertainty (on the order of $\pm$\,0.1\,dex)
in luminosities calculated at the
hottest and coolest effective temperatures estimated.

Estimated stellar parameters are listed in Table~\ref{tb:ngc55fit-pars} and
the best-fit model spectra are shown with the final reduced science spectra in
Figure~\ref{fig:ngc55spec}. The average uncertainties on the estimated stellar
parameters are significantly larger than those quoted in previous studies
using the same technique~\citep[e.g.][]{2015ApJ...805..182G}, which is owing
to the quality of the data obtained for these targets and can be illustrated
by comparing Figure~\ref{fig:ngc55spec} with Fig. 4 of~\cite{2015ApJ...805..182G}.
Despite the low S/N achieved (see Table~\ref{tb:ngc55fit-pars}),
we are still able to provide reasonable estimates of physical parameters for 14 targets, the
first spectroscopically confirmed RSGs within NGC\,55.

\subsubsection{Metallicities and spatial variations} % (fold)
\label{sub:metgrad_in_ngc55}

Assuming no spatial variations in metallicity, the average metallicity of the
sample is $-$0.48\,$\pm$\,0.13\,dex.
This value compares well to the average metallicity measured using 12 BSGs in
NGC\,55 from~\citet[$-$0.40\,$\pm$\,0.13]{2012A&A...542A..79C},
who found no evidence for spatial variations.

\cite{2016ApJ...829...70K} furthered the work on BSGs by providing metallicity
measurements for $\sim$60 BSGs in NGC\,55, increasing the number of targets
five-fold and over a larger spatial extent than
\cite{2012A&A...542A..79C}.
\cite{2016ApJ...829...70K} report a metallicity gradient of
$-$0.22\,$\pm$\,0.06\,dex/R$_{25}$,
predicting a central metallicity of $-$0.37\,$\pm$\,0.03\,dex.
Using this gradient, for the average R/R$_{25}$ of the RSG sample presented here
(R/R$_{25}\simeq 0.6$), this relationship predicts
[Z]~=~$-$0.52\,dex, in excellent agreement with the average metallicity of the current sample.

\begin{figure*}
 \centering
 \includegraphics[width=0.75\textwidth]{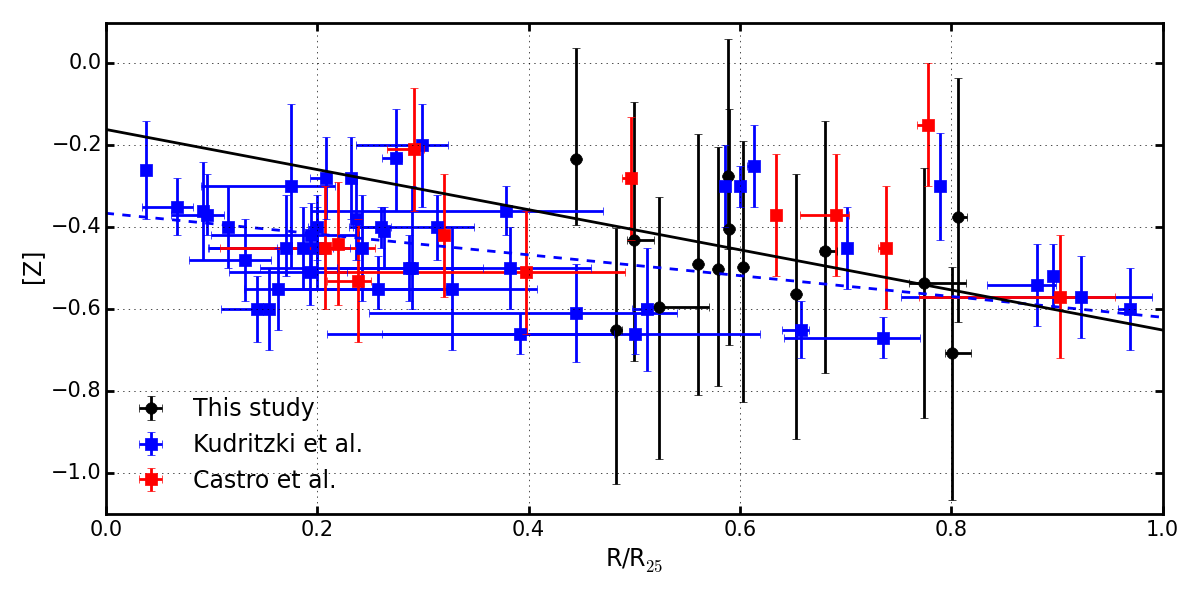}
 \caption[Metallicities for KMOS targets shown against de-projected radius]{
 Metallicities for KMOS RSGs (black points) shown against de-projected radius from the centre of NGC\,55 as defined by the Two Micron All Sky Survey~\citep{2006AJ....131.1163S} scaled by
 R$_{25}$~=~16.18\,arcmin~\citep{2004AJ....127.2031K}.
 The average metallicity of the KMOS RSGs is [Z]~=~$-$0.48\,$\pm$\,0.13\,dex.
 Results for blue supergiant stars (BSGs) from~\citet{2016ApJ...829...70K} and~\citet{2012A&A...542A..79C} are shown with red and blue squares respectively.
 The blue dashed line indicates the metallicity gradient calculated for BSGs of~\citet{2016ApJ...829...70K}, with the equation
 $y = (-0.25\,\pm\,0.07)x -(0.37\,\pm\,0.03)$.
 The black solid line indicates the metallicity gradient calculated for the RSG targets, with the equation
 $y = (-0.48\,\pm\,0.32)x -(0.16\,\pm\,0.20)$.
 The larger uncertainties on the RSG fit originate in the smaller coverage in de-projected radius and the larger uncertainties on the RSG metallicities.
 }
 \label{fig:ngc55ZvsR}
\end{figure*}

Figure~\ref{fig:ngc55ZvsR} displays the estimated metallicities shown as a
function of the de-projected radial distance from the centre of the galaxy as
defined by the Two Micron All Sky Survey
\citep[2MASS;][]{2006AJ....131.1163S}.
The RSG results are shown in black circles and results from
\cite{2016ApJ...829...70K} and
\cite{2012A&A...542A..79C} are shown in red and blue squares respectively.
% As in Figure~\ref{fig:RvsRV}, the de-projected radius is calculated by
% assuming a geometrical model with an inclination angle
% $i$~=~78\,$\pm$\,4\,\degr and a position angle $\phi$~=~109\,\degr
% \citep{1991AJ....101..447P}.
% The updated model of~\cite{2013MNRAS.434.3511W} is consistent with these values
% in inner regions of NGC\,55, where our targets are located.

Using the RSG data we find a metallicity gradient of $-$0.49\,$\pm$\,0.32\,dex/R$_{25}$
($-$0.044\,$\pm$0.029\,dex/kpc; assuming R$_{25}$~=~11.0\,kpc).
The gradient is determined using a weighted orthogonal distance
regression method ({scipy.odr}), that minimises the sum of the squared weighted
orthogonal distances from the observations to a model,
taking into account the uncertainties in the de-projected radius (abscissa) and
metallicity measurement (ordinate).

This result provides an independent measurement of the recent metallicity gradient estimated by
\cite{2016ApJ...829...70K} and predicts a central metallicity for NGC\,55 of $-$0.16\,$\pm$\,0.20\,dex in reasonable agreement with previous estimates.
The metallicity gradient estimated using the RSG data has larger uncertainties as a result of the clustered nature of the RSGs at around
$\sim$0.6\,R$_{25}$, whereas the BSG data spans a much larger galactocentric range
(as demonstrated by Figure~\ref{fig:RvsRV}).
In addition, the individual RSG metallicity measurements are also less secure than the BSG results, thus also increasing the uncertainty on the inferred gradient.

\cite{2016ApJ...829...70K} used a slightly different method to calculate the BSG metallicity gradient.
To accurately compare the results we recalculate the metallicity gradient of
\cite{2016ApJ...829...70K} with our method and find $-$0.23\,$\pm$\,0.06\,dex/R$_{25}$,
in excellent agreement with their result.

Directly comparing the results of these measurements is valid as BSGs and RSGs
are closely related in terms of stellar evolution.
These stars are predicted to be different stages in the evolution of a single
star with initial mass $>8$\,M$_{\odot}$, and as such have very similar ages.
It is important to stress that the RSG results presented here and the BSG results
\citep{2012A&A...542A..79C,2016ApJ...829...70K} are completely independent
measures of stellar metallicity using different types of stars,
different stellar model atmospheres,
different instrumentation and different wavelength regimes to make these measurements.
As such, the agreement between the two sets of results is encouraging and
further validates the use of RSGs as effective tools with which to measure extragalactic abundances.

In addition,~\cite{2017MNRAS.464..739M} analysed 25
\ion{H}{ii} regions using the so-called `direct' and `strong-line' emission line methods,
yielding abundances of various elements over a similar spatial extent to the blue and red supergiant stars in NGC\,55.
As \ion{H}{ii} regions are the birthplace of young massive stars,
their abundances should closely match that of their RSG and BSG counterparts.
Indeed, this is demonstrated in
\cite{2015ApJ...805..182G} for NGC\,300 (the companion galaxy to NGC\,55),
where the abundance gradients estimated for RSGs, BSGs and \ion{H}{ii} regions are in remarkable agreement.
However,~\cite{2017MNRAS.464..739M} found no evidence for a gradient in NGC\,55,
using the same de-projection and regression analysis as
\citet{2016ApJ...829...70K} and this study.
A similar result has previously been reported in the LMC where a metallicity
gradient has been detected in the asymptotic giant branch population of the LMC
\citep{2009A&A...506.1137C,2010MNRAS.408L..76F},
but not with abundances estimated via \ion{H}{ii} region abundances
\citep{1978MNRAS.184..569P}.

The reason for this discrepancy is, at present, unclear.
In NGC\,55, all three studies had the precision to measure this gradient
(even if somewhat ambiguously relying on solely the RSG results), and used consistent
methods to estimate this relationship.
This rules out inaccuracies in the geometric model
(for example, additional structure that is not taken into account),
as this would be systematic in all three studies.
Additional, more accurate RSG metallicities across a larger spatial extent of NGC\,55 are required
to reassess this interesting problem.

% It is interesting to note that the gradient measured using RSGs and BSGs in
% NGC\,55 is much shallower than the equivalent measurement in NGC\,300,
% the nearby companion galaxy to NGC\,55.

% subsection a_metallicity_gradient_within_ngc,55 (end)

\subsubsection{Stellar Evolution} % (fold)
\label{sub:stellar_evolution}

\begin{figure}
 \centering
 \includegraphics[width=\columnwidth]{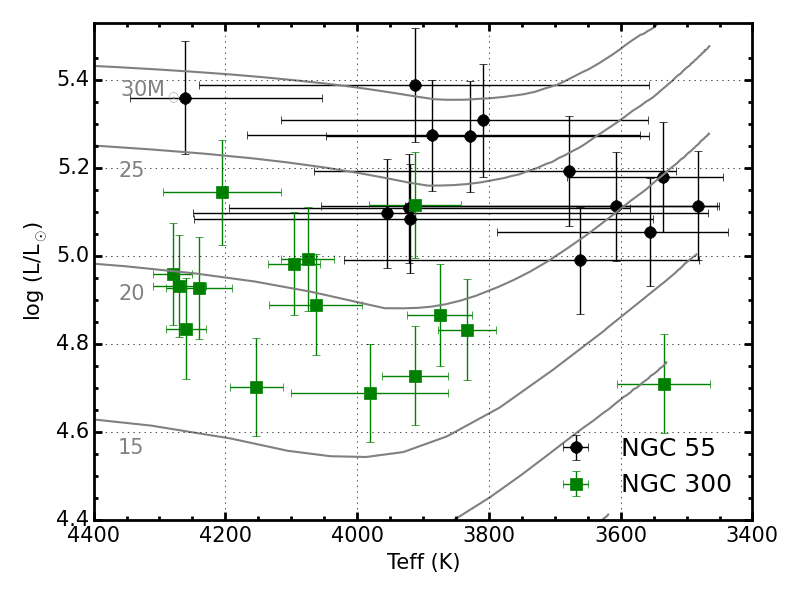}
 \caption[Hertzsprung--Russell diagram for NGC\,55]{Hertzsprung--Russell diagram for 15 RSGs in NGC\,55 shown in black circles.
 Green squares indicate results from 15 RSGs in NGC\,300 from
 \protect\cite{2015ApJ...805..182G} for comparison.
 % The best-fit linear model for each distribution is shown with the solid black and green lines.
Solid grey lines show LMC-like metallicity evolutionary models including rotation
\protect\citep{2011A&A...530A.115B}.
Results from~\cite{2015ApJ...805..182G} in NGC\,300 are shown in green for comparison.}
 \label{fig:ngc55-HRD}
\end{figure}

Luminosities are calculated using the bolometric correction of
\cite{2013ApJ...767....3D} at distance of 2.34\,$\pm$\,0.11\,Mpc.
A Hertzsprung--Russell (H-R) diagram of our targets is shown in
Figure~\ref{fig:ngc55-HRD},
using temperatures and luminosities estimated in Section~\ref{sec:ngc55results},
compared with SMC-like evolutionary tracks
\citep{2011A&A...530A.115B}.
Results for 15 RSGs in NGC\,300 are shown in green for comparison~\citep{2015ApJ...805..182G}.
By studying this figure, one notices two interesting features:
1. There appears to be a different distribution of temperatures between the sample presented here and that of NGC\,300.
2. A small systematic offset in the average luminosities of the two samples.

The distribution of temperatures of the NGC\,55 sample is skewed towards lower
temperatures where 7/15 targets have $3600 < \Teff < 3800$\,K, in contrast to 2/27 in the NGC\,300 sample.
\footnote{Note that not all of the 27 RSGs in \cite{2015ApJ...805..182G} have estimates of luminosity, and hence, some do not appear in Figure~\ref{fig:ngc55-HRD}.}
Likewise, in the NGC\,300 sample, 9/27 targets have $4200 < \Teff < 4400$\,K, whereas 1/15 NGC\,55 fall within this temperature range.
However, we caution that the uncertainties on the NGC\,55 data make a statistical comparison of the distribution of these data sets complicated.
The weighted mean and standard deviation for the two samples (NGC\,55: 3765\,$\pm$\,256\,K and NGC\,300: 4164\,$\pm$\,173\,K), highlights a $\sim 2\,\sigma$ discrepancy between the means of the distributions.

% Using a simple two-sampled Kolmogorov--Smirnov test, we can reject the null hypothesis that the two samples are drawn from the same underlying distribution
% (D~=~0.63, $p=0.0005$).
% It should be noted that this test does not take into account the uncertainties on each data point, which are significantly different between the two samples.

The differences in these distributions could be accounted for by selection effects,
as we do not select unbiased samples in these studies.
Unlike NGC\,55, NGC\,300 has a face-on orientation and the sample from~\cite{2015ApJ...805..182G} is selected over a larger range of galactocentric distance.
However, the discrepancy can likely be accounted for by the large uncertainties on the NGC\,55 data.
Additional data are required to investigate this further.

To ascertain whether or not this difference between the populations is genuine,
we revisited results for RSGs in Perseus OB-1
\citep{2014ApJ...788...58G},
the Magellanic Clouds~\citep{2015ApJ...806...21D,2016MNRAS.458.3968P} and NGC\,6822
\citep{2015ApJ...803...14P};
all of which have temperatures estimated using the same technique.
In general, these results have narrower temperature distributions than NGC\,55 with average temperatures warmer than that of NGC\,55 and cooler than in NGC\,300. The six stars in the NGC\,55 sample with $\Teff < 3700$\,K, appear to be cooler than their counterparts in the other galaxies studied.
As mentioned previously, all of the stars in the NGC\,55 sample have significant uncertainties, which make a meaningful comparison difficult.

The observed temperature distribution can be reproduced by stellar evolutionary models by varying the metallicity of the models, in general, higher metallicity models produce lower RSG temperatures.
However, the width of the temperature distribution for the NGC\,55 targets can likely be attributed to the uncertainties involved in their measurements.

A difference in the luminosity of RSG populations in a given galaxy would imply
a difference in the Humphreys--Davidson limit
\citep{1979ApJ...232..409H} between the different galaxies or a difference in the age of the stellar populations
(i.e. the most massive/youngest RSGs of the population will have already exploded as supernovae).
However, both galaxies are known to host a significant population of young and
old stars at different galactocentric radii and no evidence in the literature for a truncation of star formation in either galaxy is found.
In addition, there is no motivation for assuming differences in the Humphreys--Davidson limit.

The luminosities calculated by
\cite{2015ApJ...805..182G} are calculated using the $HST$ $F814W$ filter without correcting for the effects of extinction.
To more accurately compare luminosities of the two samples we have re-calculated their luminosities assuming
$E(B-V)=0.12$ and $R_v=3.1$
\citep{2008ApJ...681..269K}.
Figure~\ref{fig:ngc55-HRD} displays these results using this simple extinction correction.
However,~\cite{2008ApJ...681..269K} note that extinction varies significantly across NGC\,300 and can be up to $E(B-V)=0.24$.
Even if we assume that all NGC\,300 targets have $E(B-V)=0.24$ this cannot fully account for the observed offset.
The use of the $HST$ $F814W$ filter, as opposed to the $I$ filter, for the NGC\,300 sample introduces an offset of $\sim-0.02\,\log L/L_{\odot}$,
which contributes a small amount to the difference between the samples.

Another contributing factor may be that the extinction correction used for the NGC\,55 targets, $E(B-V)=0.15$,
which is calculated from the mean of the values from
\citet{2016ApJ...829...70K}, may be a slight overestimate.
This is a result of the differences in the spatial distribution of the RSG and BSG targets
i.e. the majority of BSG targets lie at a smaller galactocentric distance than the RSG studied here.
If our extinction correction is too severe, this would result in slightly overestimated luminosities for our RSG targets.

The selection criteria used here and by~\citep{2015ApJ...805..182G} both selected targets in the magnitude range $17 < I < 19$ and at the time it was assumed that both galaxies were at roughly similar distances
\citep[$\sim$1.9\,Mpc;][]{2005ApJ...628..695G,2008ApJ...672..266G}.
However, since then~\citet{2016ApJ...829...70K} have argued that NGC\,55 is at a larger distance.
The effect of this is that, in this study, we have systematically selected more luminous RSGs in NGC\,55 compared with NGC\,300.
Therefore, the fact that we have found that the RSGs in NGC\,55 are more luminous, is circumstantial evidence in support of the distance measurement of~\citet{2016ApJ...829...70K}.
In addition, when comparing the luminosities in Figure~\ref{fig:ngc55-HRD} to that of other samples of RSGs,
using the same bolometric correction
\citep[that of][]{2013ApJ...767....3D},
we find that both samples fall within the range of expected luminosities for RSGs.

In summary, selection effects and variable extinction within NGC\,300 and NGC\,55 appear to account for the small systematic offset in luminosities between the two samples.

% This small systematic difference, which we cannot account for, could be a result of uncertainties in the distance to NGC\,55
% \citep[see e.g.][]{2006A&A...455..891V,2008ApJ...672..266G,2016ApJ...829...70K}.
% For example, if the distance estimate of
% \citet[][d~=~1.93\,Mpc]{2008ApJ...672..266G} is used to calculate the luminosities of our targets,
% the systematic offset between the luminosities of the targets decreases significantly.
% In addition, when comparing the luminosities to that of other RSGs estimated using the same bolometric correction
% \citep[that of][]{2013ApJ...767....3D}, the luminosities calculated for both NGC\,55 and NGC\,300 are in good agreement with previous studies.

Interestingly, there appears to be a relationship between the temperature and
luminosity of the NGC\,55 and NGC\,300 targets, where higher temperature targets have larger luminosities.
This is in stark contrast to the relationship reported in the literature
\citet[][and see~\citealt{2013ApJ...767....3D}]{2016A&A...592A..16D},
where later spectral types correspond to more luminous stars.
In addition, this is at odds with the results of
\cite{2015ApJ...803...14P}, who find no significant correlation between luminosity
and temperature of 11 RSGs in NGC\,6822
(a Local Group dwarf irregular galaxy).
If the relationship in Figure~\ref{fig:ngc55-HRD} is real,
then one would also expect this relationship to be apparent in Figure~\ref{fig:VI}.
By analysing the observed targets in this figure, there appears to be no significant relationship between
colour and magnitude.

The explanation to observed trends in Figure~\ref{fig:ngc55-HRD}, may lie in the method used to calculate the
luminosities of these targets, as we only have a small number of photometric data points available.
As mentioned earlier, the bolometric correction used here is calibrated at a
fixed temperature.
Therefore at the most extreme temperatures in our sample, the calculated luminosities
may include a small systematic offset (on the order of $\pm$\,0.1\,dex) towards
larger (smaller) luminosities at warmer (cooler) temperatures.
To unambiguously define this relationship in NGC\,55 requires a larger sample
of RSGs with more accurate stellar parameters over a larger spatial scale.

% subsection stellar_evolution (end)

% subsection stellar_parameters (end)

% section results (end)

\section{Conclusions} % (fold)
\label{sec:ngc55conc}

We have presented KMOS spectroscopy for 18 RSGs in the Sculptor Group galaxy NGC\,55.
Radial velocities are calculated for each epoch and are shown to agree well with measurements for hot massive stars and
\ion{H}{i} gas, where all targets with reliable radial-velocity measurements are consistent with membership of NGC\,55: confirming their nature as supergiants.

Stellar parameters are estimated for 14 targets with the highest-quality observations using the $J$-band analysis technique,
which uses a grid of synthetic spectra extracted from state-of-the-art stellar model atmospheres with corrections for non-LTE in the strongest stellar absorption lines in a narrow spectral window.
The results of this study confirm the low-metallicity nature of NGC\,55 and together with the recent BSG results of
\cite{2016ApJ...829...70K}, confirm the stellar metallicity gradient.
A metallicity gradient of $-$0.48\,$\pm$\,0.32\,dex/R$_{25}$ is identified using solely the RSG data, in good agreement with recent estimates of the metallicity gradient using BSG data.
The large uncertainties on this gradient are owing to the fact that the RSGs are clustered at a radius of
$\sim$0.6\,R$_{25}$.
Interestingly, contemporary metallicity measurements in this galaxy using nebular abundances from \ion{H}{ii} regions do not follow the same trend,
potentially indicating unexpected differences between stellar and nebular abundances in NGC\,55,
which warrant further investigation.

% In contrast to current views in the literature we find that the more luminous RSGs tend to have higher temperatures, a relationship which is mirror in NGC\,300, the companion face-on spiral galaxy to NGC\,55.
% At present we have no explanation for this relationship.

Luminosities are calculated for our RSG targets and we comment on the relationship between luminosity and temperature in this galaxy, as well as the temperature distribution of the RSGs.
The estimated luminosities of RSGs in NGC\,55 are compared with those in NGC\,300 and we identify a small offset in the luminosities of these studies.
In this study, we have adopted the same selection criteria as in NGC\,300 (assuming that the two galaxies are at the same distance),
therefore, the observed offset is expected if NGC\,55 is actually at a larger distance, as reported by~\citet{2016ApJ...829...70K}.

NGC\,55 and its twin galaxy NGC\,300 are a fantastic laboratory to test theories of stellar and galactic evolution outside of the Local Group of galaxies.
This study presents a first look at metallicities estimated from RSGs in NGC\,55 using KMOS spectroscopy in order to study its massive-star population.
To determine some of the intriguing results of this initial study  more accurately we must acquire additional, higher S/N near-IR spectroscopic observations over the full spatial extent of this galaxy.

% section conclusions (end)

\section*{Acknowledgements}
The authors would like to thank the anonymous referee for helpful suggestions that have improved this publication significantly.
LRP would like to thank Ignacio Negueruela and Ross McLure for commenting on an earlier draft of this paper.
Based on observations collected at the European Organisation for Astronomical Research in the Southern Hemisphere under ESO programme 092.B-0088(A).
Support from the Ideas Plus grant of the Polish Ministry of Science and Higher Education
IdP II 2015 0002 64 is also acknowledged.

%%%%%%%%%%%%%%%%%%%%%%%%%%%%%%%%%%%%%%%%%%%%%%%%%%



\bibliographystyle{mnras}
\bibliography{journals,books} % if your bibtex file is called example.bib

% \begin{thebibliography}{99}
% \bibitem[\protect\citeauthoryear{Author}{2012}]{Author2012}
% Author A.~N., 2013, Journal of Improbable Astronomy, 1, 1
% \bibitem[\protect\citeauthoryear{Others}{2013}]{Others2013}
% Others S., 2012, Journal of Interesting Stuff, 17, 198
% \end{thebibliography}

%%%%%%%%%%%%%%%%%%%%%%%%%%%%%%%%%%%%%%%%%%%%%%%%%%

%%%%%%%%%%%%%%%%% APPENDICES %%%%%%%%%%%%%%%%%%%%%

% \appendix

% \section{Some extra material}

% If you want to present additional material which would interrupt the flow of the main paper,
% it can be placed in an Appendix which appears after the list of references.

%%%%%%%%%%%%%%%%%%%%%%%%%%%%%%%%%%%%%%%%%%%%%%%%%%
\begin{landscape}
% \scriptsize
\begin{table}
\caption[Summary of VLT-KMOS targets in NGC\,55]{Summary of VLT-KMOS targets with reliable radial-velocity estimates consistent with NGC\,55.\label{tb:n55obs-params}}
\centering
\begin{tabular}{lcccccccccl}
 \hline
 \hline
ID & $\alpha$ (J2000) & $\delta$ (J2000) & $V$ $^{\rm a}$ & $I$ $^{\rm a}$ & $F606W$ $^{\rm b}$ & $F814W$ $^{\rm b}$ & \multicolumn{2}{c}{$rv$ (\kms)} & $\langle rv\rangle$ (\kms) & Notes \\
\cline{8-9}
& &  & & & & & 14-10-2013 & 15-10-2013\\

 \hline
NGC55-RSG19 & 00:15:29.190 & $-$39:14:08.20& 19.914 & 17.731 &19.85 & 17.76 &\pp205\,$\pm$\,4\z &\pp178\,$\pm$\,7\z & 192\,$\pm$\,4\z & \\
NGC55-RSG20 & 00:15:29.520 & $-$39:15:13.00& 20.832 & 18.952 &20.86 & 19.11 &\pp194\,$\pm$\,14  &\pp220\,$\pm$\,5\z & 207\,$\pm$\,7\z & Potential blend from $HST$ image\\
NGC55-RSG24 & 00:15:31.460 & $-$39:14:46.30& 20.612 & 18.475 &20.29 & 18.38 &\pp186\,$\pm$\,6\z &\pp194\,$\pm$\,7\z & 190\,$\pm$\,5\z & \\
NGC55-RSG25 & 00:15:31.490 & $-$39:14:32.40& 20.316 & 18.394 &20.63 & 18.49 &\pp204\,$\pm$\,12  &\pp217\,$\pm$\,16  & 211\,$\pm$\,10  & \\
NGC55-RSG26 & 00:15:33.160 & $-$39:13:42.00& 20.572 & 17.964 &20.35 & 18.06 &\pp174\,$\pm$\,9\z &\pp173\,$\pm$\,8\z & 174\,$\pm$\,6\z & \\
NGC55-RSG28 & 00:15:36.160 & $-$39:15:29.40& 21.001 & 18.892 &20.87 & 18.97 &\pp233\,$\pm$\,17  &\pp161\,$\pm$\,20  & 197\,$\pm$\,13  & \\
NGC55-RSG30 & 00:15:38.030 & $-$39:14:50.20& 20.867 & 18.730 &20.79 & 18.75 &\pp212\,$\pm$\,10  &\pp215\,$\pm$\,10  & 214\,$\pm$\,7\z & \\
NGC55-RSG35 & 00:15:39.260 & $-$39:15:01.70& 20.007 & 17.872 &19.78 & 17.73 &\pp202\,$\pm$\,3\z &\pp206\,$\pm$\,4\z & 204\,$\pm$\,3\z & \\
NGC55-RSG39 & 00:15:40.260 & $-$39:15:01.00& 19.654 & 17.970 &20.36 & 18.19 &\pp206\,$\pm$\,11  &\pp192\,$\pm$\,5\z & 199\,$\pm$\,6\z & \\
NGC55-RSG43 & 00:15:40.700 & $-$39:14:50.20& 19.957 & 18.183 &20.36 & 18.25 &\pp198\,$\pm$\,6\z &\pp196\,$\pm$\,5\z & 197\,$\pm$\,4\z & Potential blend from $HST$ image\\
NGC55-RSG46 & 00:15:41.640 & $-$39:14:58.80& 21.591 & 18.441 &20.85 & 18.52 &\pp228\,$\pm$\,5\z &\pp195\,$\pm$\,6\z & 212\,$\pm$\,4\z & Potential blend from $HST$ image\\
NGC55-RSG57 & 00:15:45.590 & $-$39:15:16.40& 20.010 & 18.220 & --   & --    &\pp217\,$\pm$\,10  &\pp197\,$\pm$\,6\z & 207\,$\pm$\,6\z & \\
NGC55-RSG58 & 00:15:46.270 & $-$39:15:43.20& 20.619 & 18.400 & --   & --    &\pp236\,$\pm$\,8\z &\pp216\,$\pm$\,3\z & 226\,$\pm$\,4\z & \\
NGC55-RSG65 & 00:15:51.250 & $-$39:16:26.40& 19.706 & 17.653 & --   & --    &\pp224\,$\pm$\,5\z &\pp215\,$\pm$\,4\z & 220\,$\pm$\,3\z & \\
NGC55-RSG69 & 00:15:55.280 & $-$39:15:00.10& 20.470 & 18.666 & --   & --    &\pp231\,$\pm$\,5\z &\pp195\,$\pm$\,9\z & 213\,$\pm$\,5\z & \\
NGC55-RSG70 & 00:15:56.310 & $-$39:16:08.60& 22.300 & 18.907 & --   & --    &\pp155\,$\pm$\,12  &\pp187\,$\pm$\,9\z & 171\,$\pm$\,8\z & \\
NGC55-RSG71 & 00:15:56.900 & $-$39:15:27.50& 20.401 & 18.559 & --   & --    &\pp197\,$\pm$\,11  &\pp214\,$\pm$\,11  & 206\,$\pm$\,8\z & \\
NGC55-RSG73 & 00:15:57.710 & $-$39:15:41.50& 20.489 & 18.411 & --   & --    &\pp161\,$\pm$\,7\z &\pp178\,$\pm$\,6\z & 170\,$\pm$\,5\z & \\

% NGC55-RSG22 & 00:15:30.520 & $-$39:16:36.70& 20.406 & 18.589 & --   & --    &\pp\z95\,$\pm$\,14 &\z$-$41\,$\pm$\,26 & --             & \\
% NGC55-RSG36 & 00:15:39.520 & $-$39:16:23.10& 19.915 & 18.462 & --   & --    &$-$188\,$\pm$\,31  &$-$284\,$\pm$\,16  & --             & \\
% NGC55-RSG60 & 00:15:49.180 & $-$39:17:19.80& 21.393 & 18.847 & --   & --    &\z$-$73\,$\pm$\,39 &\pp\z26\,$\pm$\,26 & --             & \\
% NGC55-RSG67 & 00:15:53.110 & $-$39:14:13.60& 19.925 & 18.047 & --   & --    &\z\pp25\,$\pm$\,24 &\pp\z\z6\,$\pm$\,31& 175\,$\pm$\,18 & \\
\hline
\end{tabular}
\begin{itemize}
  \item [a] Ground-based data from the Araucaria Project
  \protect\citep{2006AJ....132.2556P}, with typical photometric uncertainties of 0.075 and 0.016\,mag in the $V$- and $I$-bands, respectively.
  \item [b] $HST$ ANGST photometry from
  \protect\cite{2009ApJS..183...67D}, with typical uncertainties of 0.12, 0.13\,mag in the $F606W$ and $F814W$ bands, respectively.
  % \item [c] Value excluded from average for target.
\end{itemize}
\end{table}
\end{landscape}

% Don't change these lines
\bsp	% typesetting comment
\label{lastpage}
\end{document}